# New method of image processing via statistical analysis for application in intelligent systems

Monalisa Cavalcante[,1,4], José Araújo[2,4], and José Holanda[1,2,3,4*]

[1]Programa de Pós-Graduação em Engenharia Física, Universidade Federal Rural de Pernambuco, 54518-430, Cabo de Santo Agostinho, Pernambuco, Brazil

[2]Unidade Acadêmica do Cabo de Santo Agostinho, Universidade Federal Rural de Pernambuco, 54518-430, Cabo de Santo Agostinho, Pernambuco, Brazil

[3]Programa de Pós-Graduação em Física Aplicada, Universidade Federal Rural de Pernambuco, 52171-900, Recife, Pernambuco, Brazil

[4]Group of Optoelectronics and Spintronics, Universidade Federal Rural de Pernambuco, 54518-430, Cabo de Santo Agostinho, Pernambuco, Brazil

## Abstract

Image processing has always been a topic of significant importance to society. Recently, this field has gained considerable prominence due to the development of intelligent systems. In this work, we present a new method of image processing that utilizes statistical analysis, specifically designed for applications in intelligent systems. We tested our method on a large collection of images to assess its effectiveness.

Corresponding author: *joseholanda.silvajunior@ufrpe.br

## I. Introduction

Image processing is a scientific field that focuses on the coding and analysis of images using various algorithms [1, 2]. It involves manipulating images to enhance their quality, extract information, or perform specific actions based on their content [3, 4]. This area has a wide range of applications, including medicine [5], video description [6], pattern recognition [7], and virtual reality [8], among others [11-13]. Advancements in technology and increased computational power have fueled the continuous development of techniques and algorithms in image processing [14, 15], particularly in the medical domain. Here, it is utilized for analyzing medical imaging exams, detecting diseases, and assisting in surgical procedures [16]. Additionally, image processing plays a role in security systems, such as facial recognition and the detection of suspicious movements [17].

Image processing involves several key steps, ranging from image acquisition to the extraction of relevant information [18-20]. The primary stages include acquisition, processing, segmentation, feature extraction, and classification. These techniques are crucial for ensuring that images are in optimal conditions for the subsequent steps, allowing for accurate information extraction and satisfactory results [21, 22]. Image segmentation is the process of dividing an image into regions or objects of interest. There are several techniques for segmentation, including thresholding [23, 24], region growing [25], edge detection [26], and clustering algorithms [27]. Each technique has specific characteristics and applications, making their selection dependent on the features of the images and the goals of the analysis. Feature extraction aims to identify attributes that are significant for the analysis. This can include aspects such as shape, texture, color, and intensity, among other distinctive features of the images. These characteristics are later utilized in classification, pattern recognition, indexing, and image retrieval, providing valuable information for various applications [28-31]. Image classification involves categorizing images based on their attributes and characteristics. This process is carried out using machine learning algorithms, which are trained to recognize patterns and assign labels according to predefined classes, often using a programming language. While existing approaches to image processing are important and have their advantages, they often face challenges, such as a loss of quality in the processed images and the high data density required to identify them accurately.

We have developed a new method for image processing using statistical analysis. To demonstrate the effectiveness of this method, we analyzed a large number of images. These images depict alumina membranes with nickel nanowires arranged hexagonally within their pores, obtained through scanning electron microscopy. Our images possess high resolution and excellent quality.

**II. Methodology**

Initially, we utilized ImageJ software [32] to convert all images into pixel maps. This allowed us to associate each set of pixels with their respective intensity values. We then linked these patterns to specific colors, and consequently, to their effective wavelengths. The next step involved statistically analyzing the distributions associated with the patterns of the same colors. After this phase, the statistical distributions contained all the necessary information for image identification, enabling the method to be easily implemented in intelligent systems. **Figure 1** provides a general overview of our method.

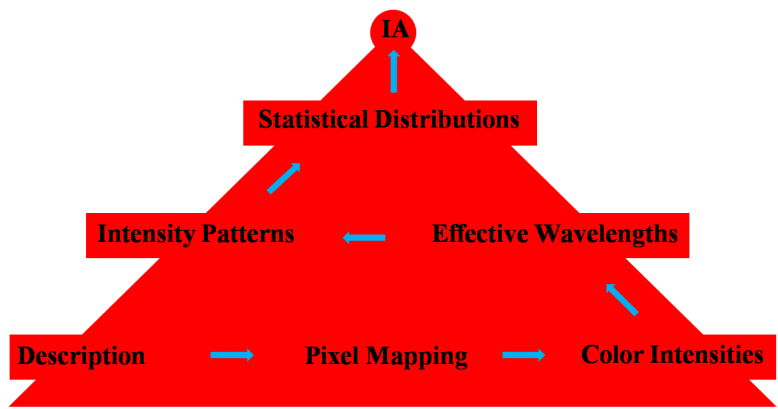

**Figure 1**. The block diagram illustrates the phases of the image processing method, which progresses through the following sequential steps: description, pixel mapping, color intensity analysis, effective wavelengths, intensity patterns, statistical distributions, and intelligent systems. This framework is essential for both the processing and analysis of images, as well as for the development of codes and data analysis.

**III. Results and discussion**

**A. Image processing**

In image processing, the primary focus is on manipulating images using computational algorithms. The quality of an image is closely linked to its pixel resolution and file size, which can significantly affect the complexity of processing tasks [28-31]. Pixel resolution is vital for determining the clarity and detail of an image, especially in applications that require high precision. However, high-resolution images often result in larger file sizes, which can strain storage and processing systems. This necessitates the use of more powerful machines and computers with greater processing capabilities. Given this context, image processing plays a critical role in streamlining computational tasks [5-13]. The processing operations include resizing, compression, and normalization, which help reduce file size without compromising visual quality.

Image processing techniques can effectively remove noise, correct distortions, and highlight important features, thereby enhancing the suitability of images for later analysis. Segmenting the image into regions of interest is another effective strategy that reduces computational load by focusing processing efforts on relevant areas [18-27]. In essence, image processing is crucial for improving both the efficiency and accuracy of analyses, even when working with high-resolution images and large file sizes, as illustrated in **Figures 2 (a) and (b)**. This approach allows for a reduction in image file size without sacrificing quality, which further facilitates the use of various tools and operations that will be described in detail below.

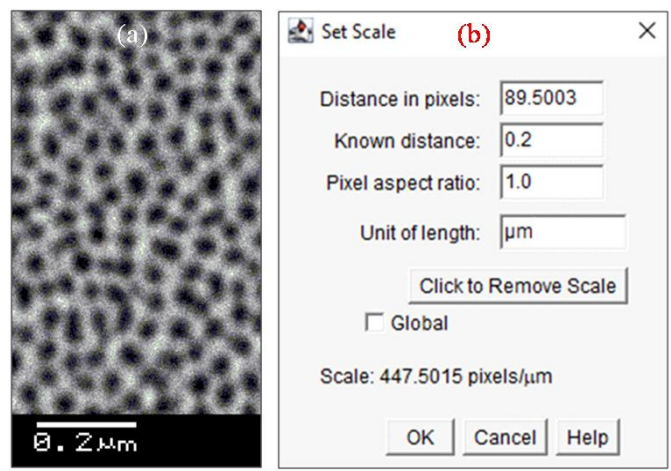

**Figure 2**: Using the ImageJ program [32], the scale was applied. **(a)** An enlarged image enhances scale visibility. **(b)** Display of the Set Scale command with the applied scale.

Using the ImageJ program [32], we performed image adjustment and processing. First, we defined an operating scale for our analysis measurements, which is provided alongside the image. This allows us to understand the dimensions we are working with. By applying the "Set Scale" command and drawing a line over the dimension bar, we can determine the number of pixels corresponding to that distance; in **Figure 2**, we considered this distance to be 0.2 with the unit of measurement in micrometers (µm). Next, we converted the image to grayscale with 8 bits. Color images, represented in RGB, contain more color channels, while grayscale images consist of only one intensity channel. For this analysis, we assumed that the dimensions of the image and the number of gray levels are integer powers of 2. When the number of gray levels equals 2, the image is classified as binary. This conversion significantly reduces computational complexity during processing, as operations like filtering and segmentation can be executed using just one channel [30-34].

Converting an image to grayscale removes color variations that may arise from differences in lighting, white balance, or environmental conditions. This simplification can enhance processing, resulting in more consistent and reliable outcomes. At this stage, we apply a smoothing filter, specifically a Bandpass Filter using Fast Fourier Transform (FFT), to clean the images. This process eliminates noise, reduces computational load, and smooths the contours, which aids in segmentation, as illustrated in **Figures 3 (a)** and **(b)**.

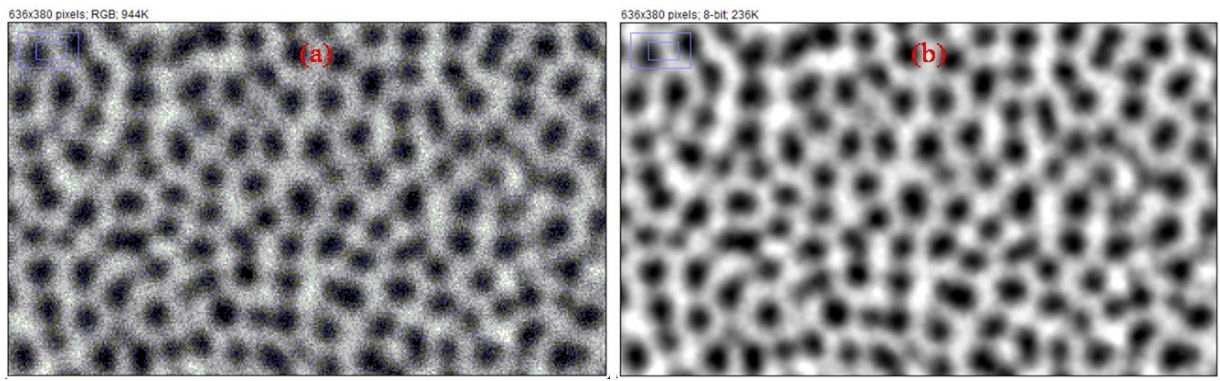

**Figure 3:** Image of electrodeposited and sectioned nickel nanowires at 200% magnification: **(a)** before conversion, and **(b)** after conversion to 8 bits and applying a Bandpass Filter.

After completing this process, the image was converted to "16 Colors" mode using the "LUT" tool. This tool changes the color of each set of pixels to a corresponding color, making it easier to identify pixel intensity, which is essential for segmentation. This is illustrated in **Figure 4**.

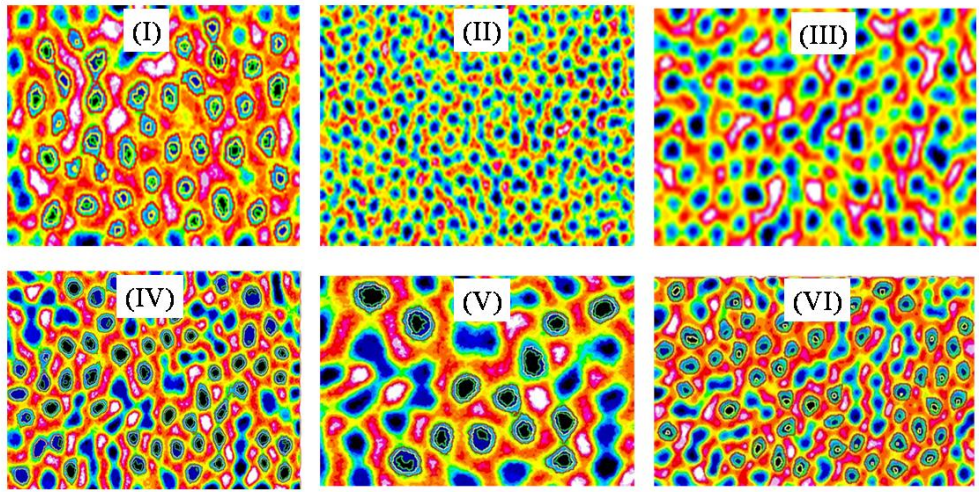

**Figure 4**: Images processed and converted to 16-color mode of electrodeposited nickel nanowires. Six regions with different contrasts are presented here (I, II, III, IV, V, and VI).

**B. Image analysis**

After image processing, the next step is analysis, which is crucial for interpreting and extracting significant information. This stage focuses on understanding the visual characteristics presented in the image, such as identifying patterns. The types of data derived from this analysis include the shape, color, texture, and position of objects within the image. Additionally, segmentation divides the image into distinct regions, while pattern detection identifies features like lines and edges. These elements can be examined through various methods, ranging from simple observation to the use of specific analysis algorithms, including statistical analysis and machine learning techniques.

The results obtained are utilized for decision-making, influenced by the methods and materials involved in the work, as well as the ability to identify patterns for more effective analysis. Additionally, image analysis often facilitates process automation, enabling tasks to be performed efficiently and accurately. The data analyzed can provide valuable insights for generating reports, predicting trends, or optimizing processes based on the specific field of application. In this context, we adopted the following steps for our samples:

1- Using the wand tracing tool in ImageJ, we outlined the regions of interest, allowing us to define the perimeter based on the grouping of pixels of the same color. This tool also includes pixels of different colors that are intertwined, as illustrated in **Figure 4**.

2- By using this approach, we can choose specific regions that are relevant for analysis. In **Figure 4**, we see the segmented elements, each with two or more defined areas. At this stage, we utilized the ROI Manager function, which stands for "Region Of Interest Manager," to separate and extract data and measurements. This tool is essential for evaluating the intensity points within each region.
3- By following this procedure, we were able to select the specific measurements that the image can provide. This allowed us to extract values related to the areas of each segment, separating them by type and associating them with each set of pixels.

**C. Analyzing data**

At this stage, we constructed histogram distribution graphs using the data obtained from image processing. Upon examining these graphs, it is evident that the area value decreases as the intensity in this region increases. It is now the responsibility of the analyzer to identify which regions exhibit the highest light intensity, which is directly related to their area, as illustrated in **Figure 5**. We calculate this intensity using the equation $I(A) = 1/A$, where A denotes the area value. This allows us to derive the distributions shown in **Figure 5** as a function of the effective wavelength. It is notable that the count and area value are inversely proportional to light intensity. This pattern is consistent across the other images analyzed. Additionally, it should be emphasized that the regions do not maintain consistent intensity ratios among themselves; each region can exhibit a wide range of intensity values, suggesting that the light may have high intensity in different areas. In these regions, we observe that the light regions often have an almost circular shape, with areas of high intensity appearing at specific points within them, not necessarily at the center.

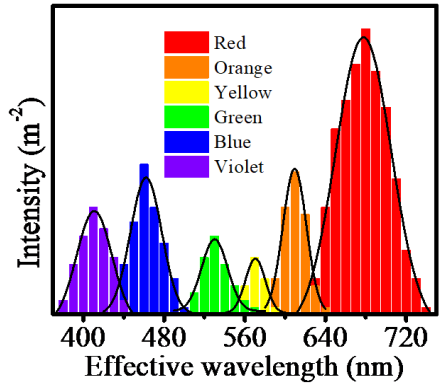

**Figure 5:** Histogram of light intensity as a function of effective wavelength. The curves represent Gaussian statistical distributions applicable in intelligent systems.

**D. Programming for image coding using intelligent systems**

The development of code for mathematical processing aims to enhance the accuracy and efficiency of data analysis through image processing in the context of artificial intelligence (AI). Our processing techniques provide a more focused and precise analysis, emphasizing the important elements that require further examination. After processing, AI can conduct both quantitative and qualitative analyses. In quantitative analysis, AI measures and quantifies specific aspects of images, such as size, shape, density, and intensity. In qualitative analysis, AI interprets the data to offer insights and diagnoses, such as detecting defects in materials or identifying medical conditions. These analyses are crucial for decision-making and for implementing corrective or preventive actions. For instance, in the manufacturing industry, early detection of defects can prevent significant issues on the production line. In healthcare, recognizing anomalies in imaging tests can lead to more accurate diagnoses and more effective treatment options.

The methodology we discuss includes a comparison step with existing databases. Artificial intelligence can compare newly processed images with historical data and previous results to identify matches, trends, and anomalies. This capability is particularly valuable in scientific research, where it is essential to validate new findings against existing data. Continuous comparison enables algorithms to learn and improve, thereby increasing the accuracy of analyses over time. For instance, in long-term studies of material behavior, the ability to compare new data with decades of previous results can uncover trends and insights that would be impossible to detect manually.

AI can generate detailed reports and graphical visualizations of analysis results. These reports may include graphs, charts, histograms, and other visual representations that help make the data and conclusions easier to understand. Automating report generation saves time and resources, enabling professionals to focus on interpreting results and implementing solutions. In a corporate environment, the ability to quickly produce detailed reports can accelerate strategic decision-making, improve responses to emerging issues, and enhance communication between teams.

**Iv. Conclusion**

In summary, we introduce a new method of image processing based on statistical analysis, which can be highly beneficial for intelligent systems. Our approach begins with the use of images that have been processed and transformed into data maps. These maps are then associated with regions of different colors, each corresponding to specific wavelengths of visible light. This method simplifies the use of images and provides a solid characterization, allowing for unique identification when implemented in intelligent systems.

**Acknowledgements**

This research was supported by Conselho Nacional de Desenvolvimento Científico e Tecnológico (CNPq) with Grant Number: 309982/2021-9, Coordenação de Aperfeiçoamento de Pessoal de Nível Superior (CAPES) with Grant Number: PROAP2024UFRPE, and Fundação de Amparo à Ciência e Tecnologia do Estado de Pernambuco (FACEPE) with Grant Number: APQ-1397-3.04/24. The authors are grateful to the Prof. Francisco Estrada from Universidad Michoacana de San Nicolas de Hidalgo for the valuable discussions on this work.

**Contributions**

J. A. and M. C. did and analyzed all the experimental measures and J. H. discussed, wrote and supervised the work.

**Conflict of interest**

The authors declare that they have no conflict of interest.

**Data Availability Statement**

Data will be made available on reasonable request.

**Referências**